\documentclass[twoside,fleqn]{ActaStyle}
\usepackage{times,amssymb,graphicx}

\newcommand{\R}{{\mathbb{R}}}
\newcommand{\half}{\frac{1}{2}}
\newcommand{\ts}{\hskip0.1ex\raisebox{-1ex}[0ex][0.8ex]%
{\rule{0.1ex}{2.75ex}\hskip0.2ex}}

\newcommand{\fig}[2]{\includegraphics[width=#1]{./figures/#2.eps}}

\newlength{\bilderlength} 
\newcommand{\bilderscale}{0.35}

\newcommand{\bilderskip}{\hspace*{0.8ex}}
\newcommand{\diagram}[1]{%
\settowidth{\bilderlength}{\bilderskip%
\includegraphics[scale=\bilderscale]{./figures/#1.eps}\bilderskip}%
\parbox{\bilderlength}{\bilderskip%
\includegraphics[scale=\bilderscale]{./figures/#1.eps}\bilderskip}}

\def\authorheading{K.J.~Wiese}

\def\shorttitle{Disorder and the functional renormalization group: A
pedagogical introduction} 

\newcommand{\rme}{{\mathrm{e}}}
\newcommand{\rmd}{{\mathrm{d}}} 
\newcommand{\nn}{\nonumber}

\newcommand{\E}{\epsilon}

\begin{document}

\pagerange{1}{11}

\title{%
DISORDERED SYSTEMS AND THE FUNCTIONAL RENORMALIZATION GROUP,\\
 A PEDAGOGICAL INTRODUCTION}

\author{
Kay J\"org Wiese\email{wiese@itp.ucsb.edu}}
{Institute of Theoretical Physics, University of
California at 
Santa Barbara\\
 Santa Barbara, CA 93106-4030, USA}

\day{April 5, 2002}

\abstract{In this article, we review basic facts about disordered
systems, especially the existence of many metastable states and and
the resulting failure of dimensional reduction. Besides techniques
based on the Gaussian variational method and replica-symmetry breaking
(RSB), the functional renormalization group (FRG) is the only general
method capable of attacking strongly disordered systems. We explain
the basic ideas of the latter method and why it is difficult to
implement. We finally review current progress for elastic manifolds in
disorder.  }



\section{Introduction}\label{intro}

Statistical mechanics is by now a rather mature branch of physics.
For pure systems like a ferromagnet, it allows to calculate so precise
details as the behavior of the specific heat on approaching the
Curie-point. We know that it diverges as a function of the distance in
temperature to the Curie-temperature, we know that this divergence has
the form of a power-law, we can calculate the exponent, and we can do
this with at least 3 digits of accuracy. This is a true success story
of statistical mechanics.  On the other hand, in nature no system is
really pure, i.e.\ without at least some disorder (``dirt'').  As
experiments (and theory) seem to suggest, a little bit of disorder
does not change the behavior much. Otherwise experiments on the
specific heat of Helium would not so extraordinarily well confirm
theoretical predictions. But what happens for strong disorder? By this
I mean that disorder completely dominates over entropy. Then already
the question: ``What is the ground-state?'' is no longer simple. This
goes hand in hand with the appearance of so-called metastable
states. States, which in energy are very close to the ground-state,
but which in configuration-space may be far apart. Any relaxational
dynamics will take an enormous time to find the correct ground-state,
and may fail altogether, as can be seen in computer-simulations as
well as in experiments. This means that our way of thinking, taught in
the treatment of pure systems, has to be adapted to account for
disorder. We will see that in contrast to pure systems, whose
universal large-scale properties can be described by very few
parameters, disordered systems demand the knowledge of the whole
disorder-distribution function (in contrast to its first few
moments). We  show how universality nevertheless emerges.

Experimental realizations of strongly disordered systems are glasses, 
or more specifically spin-glasses, vortex-glasses, electron-glasses and 
structural glasses (not treated here).
Furthermore random-field magnets, and last not least elastic systems in 
disorder. 

What is our current understanding of disordered systems? It is here
that the success story of statistical mechanics, with which I started,
comes to an end: Despite 30 years of research, we do not know much:
There are a few exact solutions, there are phenomenological methods
(like the droplet-model), and there is the mean-field approximation,
involving a method called replica-symmetry breaking (RSB). This method
is correct for infinitely connected systems, e.g.\ the SK-model
(Sherrington Kirkpatrick model), or for systems with infinitely many
components.  However it is unclear,  how far it applies
to real physical systems, in which each degree of freedom is only
coupled to a finite number of other degrees of freedom.

In this article, I report recent advances for elastic manifolds in 
random media. This system has the advantage of being approachable by 
other (analytic) methods, while still retaining all the rich physics
of strongly disordered systems.

\section{Physical realizations, model and observables}\label{model}
\begin{figure}[th]
\centerline{\parbox{6.9cm}{\fig{6.9cm}{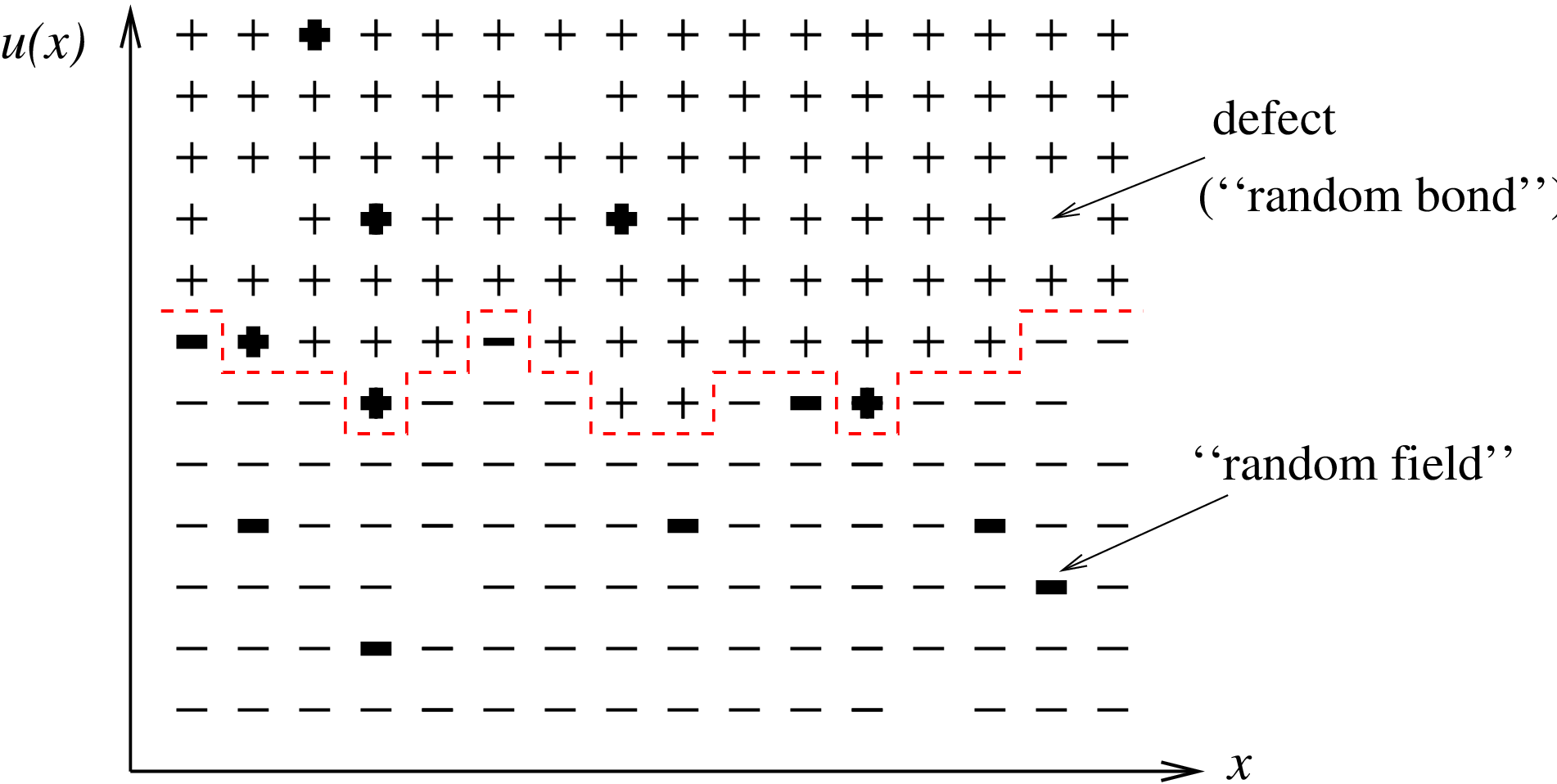}}\quad\
\parbox{6.2cm}{\fig{6.2cm}{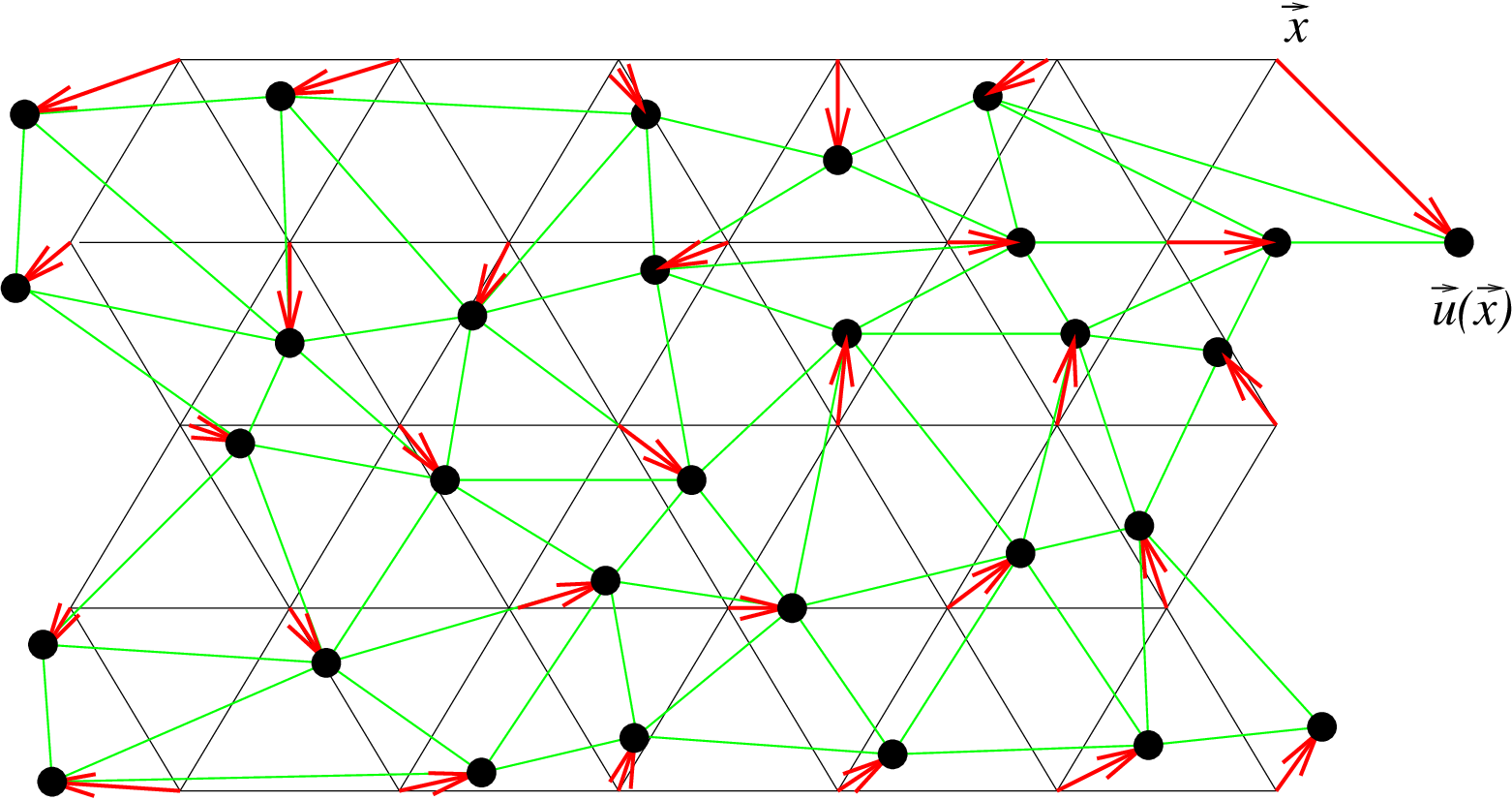}}}\smallskip

\caption{An Ising magnet at low temperatures (left) has a domain wall
described by a function
$u (x)$. Without disorder, it is flat. In the presence of disorder
it will be deformed. Right: an elastic lattice (e.g.\ vortex lattice)
deformed by disorder. This is described by a vector $\vec u
(x)$.} 
\end{figure}
Before developing the theory to treat elastic systems in a
disordered environment, let us give some physical realizations. The
simplest one is an 
Ising magnet. Imposing boundary conditions with all spins up at the
upper and all spins down at the lower boundary (see figure 1), at low
temperatures, a domain wall separates a region with spin up from a
region with spin down. In a pure system at temperature $T=0$, 
this domain wall is completely flat.  Disorder can deform the
domain wall, making it eventually rough again. Two types of disorder
are common: random 
bond (which on a course-grained level also represents missing spins)
and random field (coupling of the spins to an external random magnetic
field). Figure 1 shows, how  the domain wall is described by a displacement
field $u (x)$. 
Another example is the contact line of water (or liquid Helium),
wetting a rough substrate. (The elasticity is long range). A
realization with a 2-parameter displacement field $\vec{u} (\vec x) $
is the deformation of a vortex lattice: the position of each vortex is
deformed from $\vec x$ to $\vec x+ \vec u (\vec x)$.
A 3-dimensional example are charge density waves. 

All these models have in common, that they are described 
by a displacement field 
\begin{equation}\label{u}
x\in \R^d \ \longrightarrow\  \vec u (x) \in \R^N
\ .
\end{equation}
For simplicity, we set $N=1$ in the following.  After some initial
coarse-graining, the energy ${\cal H}={\cal H}_{\mathrm{el}}+{\cal
H}_{\mathrm{DO}}$ consists out of two parts; the elastic energy
\begin{equation}
{\cal H}_{\mathrm{el}}[u] = \int \rmd ^d x \, \half \left( \nabla u
(x)\right)^2 
\end{equation}
and the disorder
\begin{equation}
{\cal H}_{\mathrm{DO}}[u] = \int \rmd ^{d} x \, V (x,u (x))\ .
\end{equation}
In order to proceed, we need to specify the  correlations of
disorder. Suppose that fluctuations $u$ in
the transversal direction scale  as
\begin{equation}\label{roughness}
\overline{\left(u (x)-u (y) \right)^{2}}  \sim  |x-y|^{2\zeta }
\end{equation}
with a roughness-exponent $\zeta <1$. Starting from a disorder
correlator 
\begin{equation}
\overline{V (u,x)V (u',x')} = f (x-x') R (u-u')
\end{equation}
and performing one step in the RG-procedure, one has to rescale
more in the $x$-direction than in the $u$-direction. This will
eventually reduce $f (x-x')$ to a $\delta $-distribution, whereas the
structure of $R (u-u')$ remains visible. 
We therefore choose as our starting-model 
\begin{equation}
\overline{V (u,x)V (u',x')} := \delta ^{d } (x-x') R (u-u')
\ .
\end{equation}
There are a couple of useful observables. We already
mentioned the roughness-exponent $\zeta $. The second is the
renormalized (effective) disorder. It will turn out that we
actually have to keep the whole disorder distribution function $R
(u)$, in contrast to keeping a  few moments.
Other observables are higher correlation functions or the free energy.

\section{Treatment of disorder}\label{treat disorder}
Having defined our model, we can now turn to the treatment of
disorder. The problem
is to average not the partition-function, but the free energy over
disorder: $\overline{{\cal
F}}=\overline{\ln Z} $. This can be achieved by the beautiful 
 {\em replica-trick}. The idea is to write
\begin{equation}
\ln {\cal Z} = \lim_{n\to 0} \frac{1}{n}\left( \rme^{n \ln {\cal Z}}-1
\right) = \lim_{n\to 0} \frac{1}{n}\left({\cal Z}^{n}-1 \right)
\end{equation}
and to interpret ${\cal Z}^{n}$ as the partition-function of an $n$
times replicated system. Averaging $\rme ^{-\sum _{a=1}^{n}{\cal
H}_{a}}$ over disorder then leads to the {\em replica-Hamiltonian}
\begin{equation}\label{H}
{\cal H}[u] = \frac{1}{T} \sum _{a=1}^{n}\int \rmd ^{d }x\, \half
\left(\nabla u_{a} (x) \right)^{2} -\frac{1}{2 T^{2}}  \sum
_{a,b=1}^{n} \int \rmd ^{d }x\, R (u_{a} (x)-u_{b} (x))\ .
\end{equation}
Let us stress that one could equivalently pursue a dynamic
formulation. We therefore should not encounter, and in fact do not
encounter,  problems associated with the use of the replica-trick.

\section{Dimensional reduction}
There is a beautiful and rather mind-boggling theorem relating
disordered systems to pure systems 
(i.e.\ without disorder), which applies to a large class of
systems, e.g.\ random field systems and elastic manifolds in
disorder. It is called dimensional reduction and reads as
follows\cite{EfetovLarkin1977}:  

\noindent {\underline{Theorem:}} {\em A $d$-dimensional disordered
system at zero temperature is equivalent to all orders in perturbation
theory to a pure system in $d-2$ dimensions at finite temperature. }
Moreover the temperature is (up to a constant) nothing but the width
of the disorder distribution. A simple example is the 3-dimensional
random-field Ising model at zero temperature; according to the theorem
it should be equivalent to the pure 1-dimensional Ising-model at
finite temperature. But it has been shown rigorously, that the former
has an ordered phase, whereas we have all solved the latter and we
know that there is no such phase at finite temperature. So what went
wrong? Let me stress that there are no missing diagrams or any such
thing, but that the problem is more fundamental: As we will see later,
the proof makes assumptions, which are not satisfied.  Nevertheless,
the above theorem remains important since it has a devastating
consequence for all perturbative calculations in the disorder: However
clever a procedure we invent, as long as we do a perturbative
expansion, expanding the disorder in its moments, all our efforts are
futile: dimensional reduction tells us that we get a trivial and
unphysical result. Before we try to understand why this is so and how
to overcome it, let me give one more example. Dimensional reduction
allows to calculate the roughness-exponent $\zeta $ defined in
equation (\ref{roughness}).  We know (this can be inferred from
power-counting) that the width $u$ of a $d$-dimensional manifold at
finite temperature in the absence of disorder scales as $u\sim
x^{(2-d)/2}$. Making the dimensional shift implied by dimensional
reduction leads to
\begin{equation}\label{zetaDR}
\overline{\left( u (x)-u (0) \right)^{2}} \sim x^{4-d} \equiv x^{2\zeta }
\quad \mbox{i.e.}\quad \zeta =\frac{4-d}{2}\ .
\end{equation}

\section{The Larkin-length}\label{Larkin} To understand the failure of
dimensional reduction, let us turn to an interesting argument given by
Larkin \cite{Larkin1970}. He considers a piece of an elastic manifold
of size $L$. If the disorder has correlation length $r$, and
characteristic force $\bar f$, this piece will typically see a force
of strength
\begin{equation}
F_{\mathrm{DO}} = \bar f \left(\frac{L}{r} \right)^{\!\frac{d}{2}}\ . 
\end{equation}
On the other hand, there is an elastic force, which scales like 
\begin{equation}
F_{\mathrm{el}} = c\, L^{d-2}\ . 
\end{equation}
These forces are balanced at the  {\em Larkin-length} $L=L_{c}$
with 
\begin{equation}
L_{c} = \left(\frac{c^{2}}{\bar f^{2}}r^{d} \right)^{\frac{1}{4-d}}
\ .
\end{equation}
More important than this value is the observation that in all
physically interesting dimensions $d<4$, and at
scales $L>L_{c}$, the membrane is pinned by disorder; whereas on small
scales elastic energy dominates. Since the disorder has a lot of 
minima which are far apart in configurational space but close in energy
(metastability), the manifold can be in either of these minimas, and
the ground-state is no longer unique. However exactly this is assumed
in e.g.\ the proof of dimensional reduction; as is is most easily seen
in its supersymmetric formulation \cite{ParisiSourlas1979}. 

\section{The functional renormalization group (FRG)}\label{FRG}
\begin{figure}[t]
{\unitlength1mm\fboxsep0mm 
\mbox{\begin{picture} (134,30)
\put(0,25){$R (u_{a} (x)-u_{b} (x))$}
\put(5,20){$ = \diagram{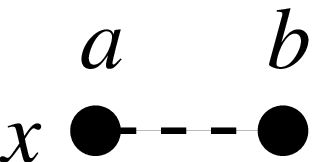}$}
\put(0,10){$C (x-y)=$}
\put(0,5) {$\ \diagram{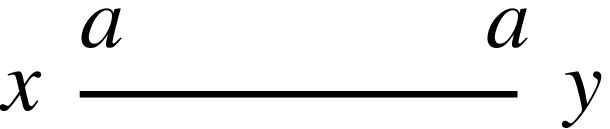}$}
\put(40,19){$\delta R (u_{a}-u_{b}) =
\diagram{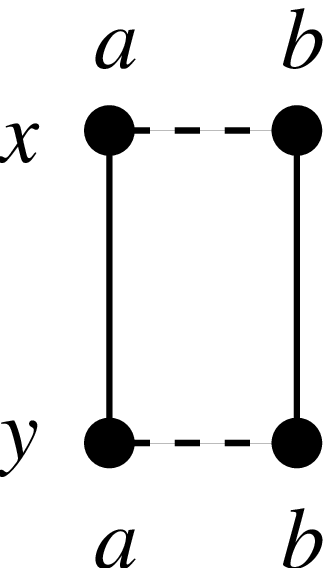}-2\diagram{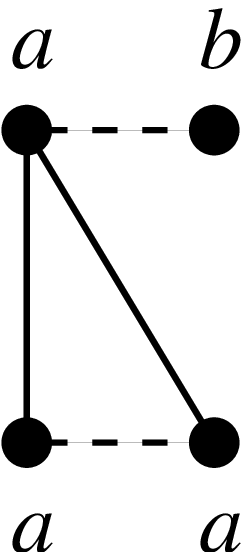} $}
\put(40,4){$=\displaystyle \int_{x-y}C (x-y)^{2} \left[  R'' (u_{a}
-u_{b})^{2} - 2 R'' (u_{a}-u_{b})R'' (u_{a}-u_{a})\right]$} 
\end{picture} }}
\caption{The disorder vertex $R (u_{a} (x)-u_{b} (x))$ and the correlation-%
function $C (x-y)$, with Fourier-transform $\tilde C
(k)=\frac{1}{k^{2}}$, which is diagonal in replica-space
(left). Contracting two disorder-vertices with two 
correlation-functions leads to the two 1-loop contributions $\delta R$
to the disorder-correlator $R$ (right). The integral $\int_{x-y}C
(x-y)^{2}=\frac{L^{\epsilon }}{\epsilon }$, where $L$ is some
IR-cutoff.} 
\end{figure}
Let us  now discuss a way out of the dilemma: Larkin's argument suggests
that $4$ is the upper critical dimension. So we would like to make an
$\epsilon =4-d$ expansion. On the other hand, dimensional reduction 
tells us that the roughness is $\zeta =\frac{4-d}{2}$ (see
(\ref{zetaDR})). Even though this 
is systematically wrong below four dimensions, it tells us correctly
that at the critical dimension $d=4$, where disorder is marginally
relevant, the field $u$ is dimensionless. This means that having
identified any relevant or marginal perturbation (as the disorder), we
find immediately another such perturbation by adding more powers of
the field. We can thus not restrict ourselves to keeping solely the
first moments of the disorder, but have to keep the whole
disorder-distribution function $R (u)$. Thus we need a {\em functional
renormalization group} treatment (FRG). 
This was first proposed in 1986 by D.\ Fisher \cite{DSFisher1986}.
Performing an 
infinitesimal renormalization, i.e.\ integrating over a momentum shell
\`a la Wilson, leads to  the flow $\partial _{\ell}
R (u)$, with ($\epsilon =4-d$)
\begin{equation}\label{1loopRG}
\partial _{\ell} R (u) = \left(\epsilon -4 \zeta  \right) R (u) +
\zeta u R' (u) + \frac{1}{2} R'' (u)^{2}-R'' (u)R'' (0)\ . 
\end{equation}
The first two terms come from the rescaling of $R$ and $u$
respectively. The last two terms are the result of the 1-loop
calculations, which are sketched in figure 2.

More important than the form of this equation is it actual solution,
sketched in figure 3.
\begin{figure}[t]
\centerline{\fig{13.4cm}{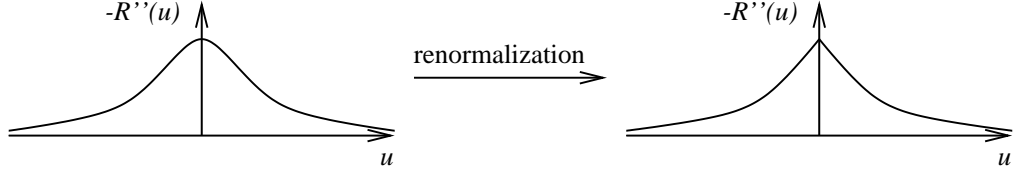}}
\caption{Change of $-R'' (u)$ under renormalization and formation of
the cusp.} 
\end{figure}
After some finite renormalization, the second derivative of the
disorder $R'' (u)$ acquires a cusp at $u=0$; the length at which this
happens is the Larkin-length. How does this overcome dimensional reduction?
To understand this, it is interesting to study the flow of the second
and forth moment. Taking a derivative of  (\ref{1loopRG}) w.r.t.\ $u$ and
setting $u$ to 0, we obtain
\begin{eqnarray}
\partial_{\ell} R'' (0) &=& \left(\epsilon -2 \zeta  \right) R'' (0) +
R''' (0)^{2} \ \longrightarrow \ \left(\epsilon -2 \zeta  \right) R''
(0)\label{R2of0}\\ 
\partial_{\ell} R'''' (0) &=& \epsilon  R'''' (0) + 3 R'''' (0)^{2} +4 R'''
(0)R''''' (0)  \ \longrightarrow\ \epsilon  R'''' (0) + 3 R''''
(0)^{2}\label{R4of0} 
\ .
\end{eqnarray}
Since $R (u)$ is an even function, $R''' (0)$ and $R''''' (0)$ are 0 and
the above equations for $R'' (0)$ and $R'''' (0)$ are in fact closed. 
Equation (\ref{R2of0}) tells us that the flow of $R'' (0)$ is
trivial and that $\zeta =\epsilon
/2\equiv \frac{4-d}{2}$. This is exactly 
the result predicted by 
dimensional reduction. The appearance of the cusp can be inferred
from equation (\ref{R4of0}). Its solution is  
\begin{equation}
R'''' (0)\ts _{\ell}= \frac{c\,\rme^ {\epsilon \ell }}{1-3\, c \left(\rme^
{\epsilon \ell} -1 \right)/ \epsilon }\ , \qquad c:= R'''' (0)\ts _{\ell=0}
\end{equation}
Thus after a finite renormalization $R'''' (0)$ becomes infinite: The cusp
appears. By analyzing the solution of the flow-equation
(\ref{1loopRG}), one also finds that beyond the
Larkin-length  $R'' (0)$ is no longer given by  (\ref{R2of0}) with $R'''
(0)^{2}=0$.  The correct interpretation  of (\ref{R2of0}),
which remains valid after the cusp-formation, is (for details see below)
\begin{equation}
\partial_{\ell} R'' (0) = \left(\epsilon -2 \zeta  \right) R'' (0)  +R''' (0^{+})^{2} \label{R2of0after}\ .
\end{equation} 
Renormalization of the whole function thus overcomes
dimensional reduction.
The appearance of the cusp also explains why  dimensional reduction
breaks down. The simplest way to see 
this is by redoing the proof for elastic manifolds in disorder, which
in the absence of disorder is a simple Gaussian theory. Terms
contributing to the 2-point function involve $R'' (0)$, $TR'''' (0)$
and higher derivatives of $R (u)$ at $u=0$, which all come with higher powers
of $T$. To obtain the limit of $T\to 0$, one sets $T=0$,
and only $R'' (0)$ remains. This is the dimensional reduction
result. However we just saw that $R'''' (0)$ becomes infinite. Thus
$R'''' (0) T$ may also contribute, and the proof fails.

\section{Why is a cusp necessary?}
The appearance of a cusp might suggest that our approach is fatally
ill. Let me present a simple argument, why a cusp {\em  is a 
physical necessity and not an artifact.} To this aim, consider a
toy model with only one Fourier-mode $u=u_{q}$
\begin{equation}\label{toy}
{\cal H}[u] = \half q^{2 } u^{2} + \sqrt{\epsilon }\, \tilde V (u)
\ .
\end{equation}
Since equation (\ref{1loopRG}) has a fixed point of order $R (u)\sim
\epsilon $ for all $\epsilon >0$, $V (u)$ scales like $\sqrt{\epsilon
}$ for $\epsilon $ small and we have made this dependence explicit in
(\ref{toy}) by using $V (u)= \sqrt{\epsilon }\tilde V (u)$. The only
further input comes from the physics: For $L<L_{c}$, i.e.\ before we
reach the Larkin length, there is only one minimum, as depicted in
figure 4. On the other hand, for $L>L_{c}$, there are several
minima. Thus there is at least one point for which
\begin{equation}
\frac{\rmd ^2}{\rmd u^2}\, {\cal H}[u] = q^{2 } + \sqrt{\epsilon }\, \tilde V''
(u) < 0
\ .
\end{equation}
In the limit of $\epsilon \to 0$, this is possible if and only if
$\frac{1}{\epsilon }R'''' (0)$, which a priori should be finite for
$\epsilon \to 0$, becomes infinite:
\begin{equation}
\frac{1}{\epsilon }R'''' (0) = \overline{V'' (u)V'' (u')}\ts _{u=u'} = \infty 
\ .
\end{equation}
This argument shows that a cusp is indeed a physical necessity.
\begin{figure}[t]{\unitlength1mm\fboxsep0mm 
\mbox{\begin{picture} (134,20)
\put(17,16.5){${\cal H}[u]$}
\put(52,1){$u$}
\put(98,16.5){${\cal H}[u]$}
\put(60,13.7){$L<L_{c}$} 
\put(61,5.5){$L>L_{c}$} 
\put(132.5,1){$u$}
\put(0,1){\fig{132mm}{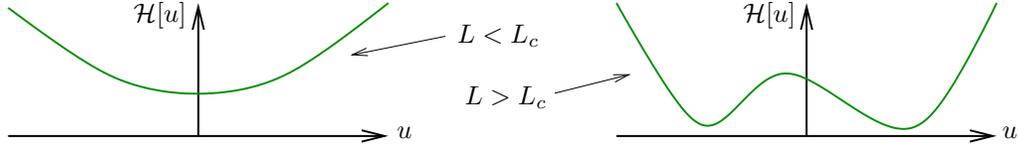} } 
\end{picture} }}
\caption{The toy model (\ref{toy}) before (left) and after (right) the
Larkin-scale.}
\end{figure}

\section{Beyond 1 loop?}\label{beyond1loop}
Functional renormalization has successfully been
applied to a bunch of problems at 1-loop order. From a field theory,
we  however demand more. Namely that it\medskip

$\bullet$ allows for systematic corrections beyond 1-loop order\smallskip

$\bullet$ be renormalizable\smallskip

$\bullet$ and thus allows to make universal predictions.\medskip

\noindent 
However, this has been a puzzle since 1986, and it has even been
suggested  that the theory is not renormalizable due to the
appearances of terms of order $\epsilon ^{\frac{3}{2}}$. Why is the
next order so complicated? The reason is that it  involves
terms proportional to $R''' (0)$. A look at figure 3 explains the
puzzle. Shall we use the symmetry of $R (u)$ to conclude that $R'''
(0)$ is 0? Or shall we take the left-hand or right-hand derivatives,
related by
\begin{equation}
R''' (0^{+}) := \lim_{{u>0}\atop {u\to 0}} R ''' (u) = -
\lim_{{u<0}\atop {u\to 0}} R ''' (u) =:- R''' (0^{-}) .
\end{equation}
In the following, I will present the solution of this puzzle. First at
2-loop order \cite{ChauveLeDoussalWiese2000a} and then at large $N$
\cite{LeDoussalWiese2001}. The latter approach allows for another
independent control-parameter, and sheds further light on the cusp-formation.

\section{Results at 2-loop order}\label{2loop}
For the flow-equation at 2-loop order, we find
\cite{ChauveLeDoussalWiese2000a}  
\begin{eqnarray}\label{2loopRG}
\partial _{\ell} R (u) &=& \left(\epsilon -4 \zeta  \right) R (u) +
\zeta u R' (u) + \frac{1}{2} R'' (u)^{2}-R'' (u)R'' (0) \nn \\
&& + \frac{1}{2}\left(R'' (u)-R'' (0) \right)R'''
(u)^{2}-\frac{1}{2}R''' (0^{+})^{2 } R'' (u) \ .
\end{eqnarray}
The first line is the result at 1-loop order, already given in
(\ref{1loopRG}). The second line is new. The most interesting term is
the last one, which involves $R''' (0^{+})^{2}$ and which we therefore
call {\em anomalous}.  The hard task is to fix the prefactor
$(-\frac{1}{2})$.  We have up to now invented six algorithms
to fix it; one leads to inconsistencies and shall not be reported
here. The other five algorithms are consistent with each other: The
sloop-algorithm, recursive construction, reparametrization invariance,
renormalizability, and potentiality. For lack of space, we restrain
our discussion to the last two 
ones. At 2-loop order  the following diagram appears
\begin{equation}\label{rebi}
\diagram{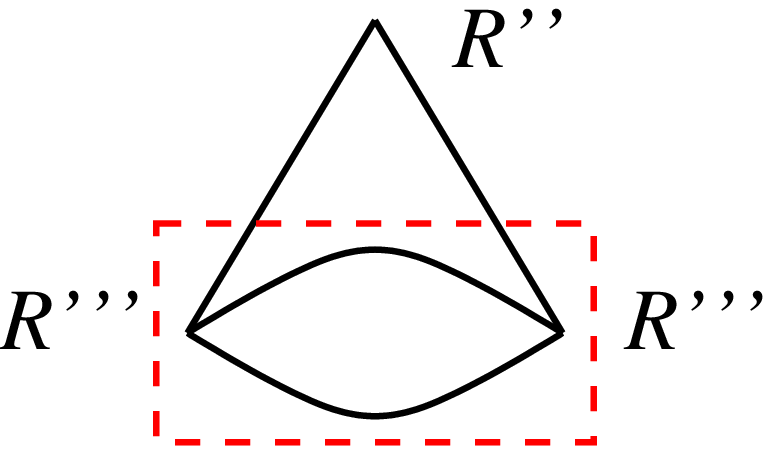}\  \longrightarrow\ \frac{1}{2}\left(R'' (u)-R'' (0)
\right)R''' 
(u)^{2} -\half R'' (u)R''' (0^{+})^{2}
\end{equation}
leading to the anomalous term. The integral (not written here) contains a
subdivergence, which is indicated by the box. Renormalizability
demands that its leading divergence (which is of order $1/\epsilon
^{2}$)  be canceled by a 1-loop counter-term. The latter is
unique thus fixing the prefactor of the anomalous term. (The idea is
to take the 1-loop correction $\delta R$ in figure 2 and replace one
of the $R''$ in it by $\delta R''$ itself, which the reader can check
 to leading to the terms given in (\ref{rebi}) plus terms
which only involve even derivatives.)

Another very physical demand is that the problem remain potential,
i.e.\ that forces still derive from a potential. The force-force
correlation function being $-R'' (u)$, this means that the flow of
$R' (0)$ has to be strictly 0. (The simplest way to see this is to
study a periodic potential.) From (\ref{2loop}) one can check that
this does not remain true if one changes the prefactor of the last
term in (\ref{2loop}); thus fixing it.

Let us give some results for cases of physical interest. First
of all, in the case of a periodic potential, which is relevant for
charge-density waves, the fixed-point function
can be calculated analytically as  (we choose period 1, the following
is for $u\in \left[0,1 \right]$)
\begin{equation}
R^{*} (u) = - \left(\frac{\epsilon }{72}+\frac{\epsilon ^{2}}{108}+O
(\epsilon ^{3}) \right) u^{2} (1-u)^{2} +\mbox{const.}
\end{equation}
This leads to a universal amplitude. 
In the case of random field disorder (short-ranged force-force
correlation function) $\zeta =\frac{\epsilon }{3}$. For random-bond
disorder (short-ranged potential-potential correlation function) we
find numerically
$\zeta = 0.208 298 04 \epsilon +0.006858 \epsilon ^{2}$. This compares
well with numerical simulations, see figure 5.

\begin{figure}\leftline{\small
\begin{tabular}{|c|c|c|c|c|}
\hline
$\zeta _{\rm}$ & one loop & two loop & estimate & 
simulation and exact\\
\hline
\hline
$d=3$  & 0.208 &  0.215  & $0.215\pm 0.01$  & 
$0.22\pm 0.01$ \cite{Middleton1995}  \\
\hline
$d=2$ &0.417 &0.444 &$0.42\pm 0.02$ &  $0.41\pm 0.01$ \cite{Middleton1995} \\
\hline
$d=1$ & 0.625 & 0.687 &  $0.67\pm 0.02$ & $2/3$ \\
\hline
\end{tabular}}\medskip 
\caption{Results for $\zeta $ in the random bond case.}
\end{figure}

\section{Large $N$}\label{largeN}
In the last section, we have discussed renormalization in a loop
expansion, i.e.\ expansion in $\E$. In order to independently check
consistency it is  good to have another non-perturbative
approach. This is achieved by the large-$N$ limit, which can be solved
analytically and to which we turn now. We start from 
\begin{eqnarray}\label{HlargeN}
{\cal H}[\vec u,\vec j ] &=& \frac{1}{2T} \sum _{a=1}^{n}\int_{x} 
 \vec u_{a} (x)\left(-\nabla^{2}{+}m^{2} \right) \vec u_{a} (x) - \sum
_{a=1}^{n}\int_{x} \vec{j}_{a} (x)\vec{u}_{a} (x)  \nn \\
&&   -\frac{1}{2 T^{2}}  \sum
_{a,b=1}^{n} \int_x B \left((\vec u_{a} (x)-\vec u_{b} (x))^{2} \right)\ .
\end{eqnarray}
where in contrast to (\ref{H}), we use an $N$-component field
$\vec{u} $. For $N=1$, we identify $B (u^{2} )=R (u)$. We also
have added a mass $m$ to regularize the theory in the infra-red and a
source $\vec{j} $ 
to calculate the effective action $\Gamma (\vec u) $ via a Legendre
transform. For large $N$ 
the saddle point equation reads 
\begin{equation}\label{saddlepointequation}
\tilde B' (u_{ab}^{2}) = B' \left(u_{ab}^{2}+2 T I_{1} + 4
I_{2} [\tilde B' (u_{ab}^{2})-\tilde B' (0)] \right)
\end{equation}
This equation gives the derivative of the effective (renormalized)
disorder $\tilde B$ as 
a function of the (constant) background field $u_{ab}^{2}= (u_{a}-u_{b})^{2}$
in terms of: the derivative of the microscopic (bare) disorder $B$,
the temperature $T$ 
and the integrals $I_{n}:= \int_{k}\frac{1}{\left(k^{2}+m^{2}
\right)^{n}}$.  

The saddle-point equation can again be turned into a closed functional
renormalization group equation for $\tilde B$ by taking the derivative
w.r.t.\ $m$:
\begin{equation}\hspace{-0.9 cm}
\partial _{l}\tilde B (x)\equiv  -\frac{m \partial }{\partial m}\tilde B (x)
=\left(\epsilon {-}4\zeta  \right)\! \tilde B (x) + 2 \zeta x \tilde B'
(x)+\frac{1}{2}\tilde B' (x)^{2}-\tilde B' (x) \tilde B' (0)+  \frac{\epsilon T
\tilde B' (x)}{\epsilon {+}\tilde B'' (0)}\,\,\,
\end{equation}
This is a complicated nonlinear partial differential
equation. It is therefore surprising, that one can find  an analytic
solution. (The trick is to 
write down the flow-equation for the inverse function of  $\tilde B'
(x)$, which is linear.) Let us only give the results of this analytic
solution: First 
of all, for long-range correlated disorder of the form $\tilde B'
(x)\sim x^{-\gamma }$, the exponent $\zeta $ can be calculated
analytically as 
$\zeta =\frac{\epsilon }{2 (1+\gamma )}\ . $
It agrees with the replica-treatment in \cite{MezardParisi1991} and the
1-loop treatment in 
\cite{BalentsDSFisher1993}.
Second, it demonstrates that  before the Larkin-length,
$\tilde B (x)$ is analytic and thus  dimensional reduction holds. Beyond the
Larkin length, $\tilde B'' (0)=\infty $, a cusp appears and
dimensional reduction is incorrect. This shows again that the cusp is not an
artifact of the perturbative expansion, but an important property even
of the exact solution of the problem (here in the limit of large
$N$). 

\section{Relation to Replica Symmetry Breaking (RSB)}\label{RSB}
\begin{figure}
\centerline{\fig{8cm}{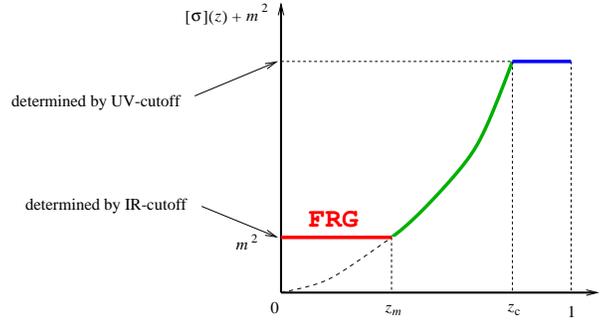}}
\caption{The function $\left[\sigma \right] (u)+m^{2}$ as given in
\cite{MezardParisi1991}.} \vspace{-0.1cm}
\end{figure}
There is another treatment of the limit of large $N$ given by M\'ezard and
Parisi \cite{MezardParisi1991}. They start from (\ref{HlargeN})  
but {\em without}\/ a source-term $j$. In the limit of large $N$, a
Gaussian variational ansatz of the form 
\begin{eqnarray}\label{HlargeNMP}
{\cal H}_{\mathrm g}[\vec u] &=& \frac{1}{2T} \sum _{a=1}^{n}\int_{x} 
 \vec u_{a} (x)\left(-\nabla^{2}{+}m^{2} \right) \vec u_{a} (x) 
   -\frac{1}{2 T^{2}}  \sum
_{a,b=1}^{n} \sigma_{ab} \, \vec u_{a} (x)\vec u_{b} (x)
\end{eqnarray}
becomes exact. The art is to make an appropriate ansatz for
$\sigma_{ab}$. The simplest possibility, $\sigma _{ab}=\sigma $ for
all $a\neq b$ reproduces the dimensional reduction result,  which
breaks down at the Larkin length. Beyond that scale, 
a replica symmetry broken (RSB) ansatz for $\sigma
_{ab}$ is suggestive. To this aim, one can break $\sigma _{ab} $ into four
blocks of equal size, choose one value for the both outer diagonal
blocks, and then iterate the procedure on the diagonal blocks. One
finds that the more often one iterates, the better the
results. In fact, one has to repeat this procedure 
infinite many times. This seems like a hopeless endeavor, but Parisi
has shown that the infinitely often replica symmetry broken matrix can
be parameterized by a function $[\sigma] (z)$ with $z\in \left[0,1
\right]$. In the SK-model, $z$ has the interpretation of an overlap
between  replicas. While there is no such simple interpretation for
the model (\ref{HlargeNMP}), we retain that $z=0$ describes distant
states, whereas $z=1$ describes  nearby states. The solution of the
large-$N$ saddle-point equations leads to the curve depicted in figure 6. 
Knowing it, the 2-point function is given by
$\left< u_{k}u_{-k} \right>=\frac{1}{k^{2}}\left(1+\int_{0}^{1} \frac{\rmd
z}{z^{2}}  \frac{\left[\sigma  \right] (z)+m^{2}}{k^{2}+\left[\sigma  \right] (z)+m^{2}} \right)$.
The important question is: What is the relation between the
two approaches, which  both pretend to calculate the same
2-point function?   
Comparing the analytical solutions, we find that the 2-point
function given by FRG is the same as that of RSB, if in the latter
expression we only take into account the contribution from the most
distant states, i.e.\ those for $z$ between 0 and $z_{m}$ (see figure
6). To understand why this is so,  we have to remember that the two
calculations 
were done under quite different assumptions: In contrast to the
RSB-calculation, the FRG-approach calculated the partition function
in presence of an external field $j$, which was then used to give via
a Legendre transformation the effective action. Even if the field $j$ is
finally turned to 0, the system might remember its preparation, as
is the case for a magnet: Preparing the system in presence of a
magnetic field will result in a magnetization which aligns with this
field. The magnetization will remain, even if finally the field is
turned off. The same phenomena happens here: By explicitly breaking
the replica-symmetry through an applied field, all replicas will
settle in  distant states, and the close states from the
Parisi-function $\left[\sigma \right] (z)+m^{2}$ (which describes {\em
spontaneous} RSB) will not contribute. 
However the full RSB-result can be
reconstructed by remarking that the part of the curve between $z_{m}$
and $z_{c}$ is independent of the infrared cutoff
$m$, and then integrating over $m$ \cite{LeDoussalWiese2001}.  
We also note that a similar effective action  has been proposed in
\cite{BalentsBouchaudMezard1996}. While it agrees qualitatively, it
does not reproduce the correct FRG 2-point function, as
it should.

\section{Perspectives}\label{perspectives} More interesting problems
have been treated by the above methods, and much more has to be
done. Besides equilibrium problems,
driven systems are also studied experimentally. An example is the
domain wall in a random field magnet, driven through the system by an
applied magnetic field.  This was treated in
\cite{NattermanStepanowTangLeschhorn1992,NarayanDSFisher1993a}, and it
was concluded that for non-periodic disorder, there is only one fixed
point, describing both random bond and random field disorder. Our
2-loop calculations
\cite{ChauveLeDoussalWiese2000a,LeDoussalWieseChauve2002} present the
first consistent field theory, capable to distinguish between statics
and dynamics. They also
show that a conjecture by \cite{NarayanDSFisher1993a} that
$\zeta=\frac{\epsilon }{3}$ be exact to all orders is violated at
second order. 
We have applied the same
methods to the statics at 3-loop order and to the random field problem.
An expansion in $1/N$, (by now we have
obtained the effective action), should allow to finally describe such
notorious problems as the strong-coupling phase of the
Kardar-Parisi-Zhang equation. Finally, it is still open of whether FRG
can also be applied to spin-glasses as e.g.\ the SK-model. We leave
this problem as a challenge to the reader.

\begin{ack}
It is a pleasure to thank the organizers of RG 2002 for the
opportunity to give this lecture. I am grateful
to Arun Paramekanti for a 
critical reading of the manuscript, and  to  my 
collaborators  Pierre Le Doussal and Pascal
Chauve for all their enthousiasm and dedicated work. 
\end{ack}

\end{document}